\documentclass[prd,preprintnumbers,twocolumn,nofootinbib,aps]{revtex4}
\usepackage{graphicx}
\usepackage{dcolumn}
\usepackage{bm}
\usepackage{epsfig,amsmath}
\usepackage{amssymb}
\usepackage{subfigure}
\usepackage{float}
\usepackage{ulem}
\usepackage[usenames,dvipsnames]{color}
\usepackage[pagebackref=false, colorlinks=true]{hyperref}
\definecolor{redish}{rgb}{0.7,0.2,0.0}  
\definecolor{bluish}{rgb}{0.2,0.5,0.8}

\hypersetup{linkcolor=redish,          
                  citecolor=blue,        
                  filecolor=magenta,      
                  urlcolor=bluish}          

\DeclareFontFamily{U}{rsfs}{}         
\DeclareFontShape{U}{rsfs}{m}{n}{<5> rsfs5 <6><7> rsfs7          %
  <8><9><10><10.95><12><14.4><17.28><20.74><24.88> rsfs10}{}     %
\DeclareMathAlphabet{\mathfs}{U}{rsfs}{m}{n}

\def \f{\frac}

\def \o{\omega}

\def \a{\alpha}
\def \b{\beta}

\def \s{\sigma}

\def \p{\partial}

\def \a{\alpha}

\begin{document}
\title{Primordial Black Holes having Gravitomagnetic Monopole}

\author{Chandrachur Chakraborty}
\email{chandrachur.c@manipal.edu} 
\affiliation{Manipal Centre for Natural Sciences, Manipal Academy of Higher Education, Manipal 576104, India}

\author{Sudip Bhattacharyya}
\email{sudip@tifr.res.in}
\affiliation{Department of Astronomy and Astrophysics,
Tata Institute of Fundamental Research, Mumbai 400005, India}

\begin{abstract}
A primordial black hole (PBH) is thought to be made of the regular matter or ordinary mass ($M$) only, and hence could have already been decayed due to the Hawking radiation if its initial ordinary mass were $\lesssim 5 \times 10^{11}$ kg. Here, we study the role of gravitomagnetic monopole for the evaporation of PBHs, and propose that the lower energy PBHs (equivalent to ordinary mass $M << 5\times 10^{11}$ kg) could still exist in our present Universe, if it has gravitomagnetic monopole. If a PBH was initially made of both regular matter and gravitomagnetic monopole, the regular matter could decay away due to the Hawking radiation. The remnant gravitomagnetic monopole might not entirely decay, which could still be found as a PBH in the form of the pseudo `mass-energy'. If a PBH with $M \gtrsim 5 \times 10^{11}$ kg is detected, one may not be able to conclude if it has gravitomagnetic monopole. But, a plausible detection of a relatively low energy (equivalent to $2.176 \times 10^{-8}$ kg $< M \lesssim 5\times10^{11}$ kg) PBH in future may imply the existence of a gravitomagnetic monopole PBH, which may or may not contain the ordinary mass.
\end{abstract}

\maketitle 

\section{Introduction\label{s1}}
From the classical field equation point of view, the matter is classified into the two categories:  gravitoelectric matter (regular matter or ordinary mass) and gravitomagnetic matter (gravitomagnetic monopole or magnetic mass) \cite{shen02, shen04}.
Due to their different gravitational features, the concept of ordinary mass does not have any significance for the gravitomagnetic matter \cite{shen02}. 
The ordinary mass $M$ and gravitomagnetic monopole (GMM) $n$ are usually defined by the Komar formula (in $G=c=1$ unit) as \cite{rs, str}
\begin{eqnarray} 
 M &\equiv & \f{1}{8\pi}\int_{S_{\infty}^2} *dk 
\end{eqnarray}
and,
\begin{eqnarray} 
 n & \equiv & \f{1}{8\pi}\int_{S_{\infty}^2} dk 
\end{eqnarray}
respectively, where, $k=(\p_t)^b=g_{0\mu}dx^{\mu}$ is the one-form of the timelike Killing vector $(\p_t)$ and the integral is taken over a spacelike 2-surface ($S$) at spacelike infinity. $*$ is the well-known Hodge dual.
As the gravitomagnetic charge is the duality of gravitoelectric charge (ordinary mass), it is also termed as the `dual mass' as well as NUT (Newman-Unti-Tamburino) charge \footnote{Although the NUT charge is called in different names, such as, gravitomagnetic matter, gravitomagnetic monopole, gravitomagnetic charge, dual mass or magnetic mass, it is needless to say that all these are the same thing. Similarly, it should also be noted that the gravitoelectric matter, regular matter, ordinary mass or gravitoelectric charge are also the same thing.} \cite{nut}. The source of NUT parameter $(n)$ is the angular momentum $2n$ unit per unit length uniformly distributed along an axis \cite{bon}. Bonnor \cite{bon} physically interpreted this NUT charge as `a linear source of pure angular momentum' \cite{dow,rs}. This specific angular momentum is not like the intrinsic angular momentum of a rigid body, or a Kerr black hole (BH). For instance, this specific angular momentum may not decrease due to the Penrose process, as there is no ergosphere at all in the absence of intrinsic spin/angular momentum \cite{hel}.

If the Schwarzschild spacetime contains the GMM aka NUT charge, it is considered as the Taub-NUT spacetime. The Taub-NUT spacetime is described by the two parameters: ordinary mass $(M)$ and GMM $(n)$ or NUT charge. The NUT spacetime is a vacuum solution of the Einstein equation. Here the vacuum is defined by the vanishing of the symmetric part of the Einstein tensor. The antisymmetric part of the Einstein tensor of the NUT spacetime does not
vanish along the axisymmetric pole \cite{gcyl}. This makes the Taub-NUT spacetime asymptotically non-flat, despite the fact that the Riemann tensor vanishes ($R_{\mu\nu\a\b} = \mathcal{O}\left( 1/r^3
\right)$) for $r \rightarrow \infty$ as the Schwarzschild case \cite{mis}. The non-vanishing $g_{t\phi}$ term leads to the anisotropy at $r \rightarrow \infty$ due to the presence of $n$. This is the reason that the Taub-NUT metric describes a homogeneous and anisotropic model of the Universe \cite{mis}. As the Universe is considered homogeneous and isotropic, other astrophysical objects may also contain the GMM to reduce or nullify its effect \cite{gcyl} to make the Universe `homogeneous and isotropic'. Another intriguing feature of the Taub-NUT spacetime is that it includes the string singularity \cite{rs2}.
To avoid this string singularity, Misner \cite{mis} imposed the time periodicity, which raises the causality violation issue \cite{rs2} due to the presence of the closed-timelike curve (CTC).
It has recently been shown \cite{cle1, cle2} that the Taub-NUT spacetime is free from causal pathologies for freely falling observers if the time periodicity is not imposed, and hence, some longstanding obstructions to
accept the Taub-NUT solution as physically relevant are removed \cite{cle1}. It is also possible to formulate \cite{hen, bor1} a consistent thermodynamics of Taub-NUT spacetime in the presence of Misner string. Regarding CTC, one can also apply the Novikov self-consistency principle/conjecture (only self-consistent trips back in time would be permitted \cite{nov,fn})  to avoid the causality violation issue \cite{gcyl} for the phenomena occurring in the Taub-NUT black holes. 

It was suggested in \cite{lnbl} that the signatures of GMM might be found in the spectra of supernovae, quasars,
or active galactic nuclei (see also \cite{kag,liu, cc}). In a recent paper, the first observational indication of GMM has been reported \cite{cbgm} based on the X-ray observations of an
accreting astrophysical collapsed object  (GRO J1655--40), which could be better described by the Kerr-Taub-NUT spacetime \cite{cbgm2} instead of the Kerr spacetime. In another work \cite{gcyl}, it has been shown that the observational constraints  (proposed by the Event Horizon Telescope collaboration) on the size and circularity of the
M87* shadow do not exclude the possibility that this collapsed object can contain GMM.  
In this paper, we study the role of such a GMM for the evaporation of PBHs.

The paper is organized as follows. In Sec. \ref{s2}, we discuss the possible formation scenario of the Taub-NUT PBHs. Sec. \ref{s3} and Sec. \ref{s4} study the evaporation of  spinning PBHs and Taub-NUT PBHs, respectively. We conclude in Sec. \ref{s5}.

\section{Possible formation of Taub-NUT PBHs\label{s2}} 
A GMM has a  gravitational effect \cite{shenp}, as its energy is non-zero. For instance, a wire without ordinary mass and carrying electric current can also show the gravitational effect, as the energy of the magnetic field generated by the current contributes to the gravitational (ordinary) mass of the system \cite{bon60}. Moreover, given that the vacuum is a perfect fluid, the gravitoelectric field (i.e., the Newtonian field of gravity) produced by
the gravitoelectric charge (ordinary mass) of the vacuum quantum fluctuations is exactly
cancelled by the gravitoelectric field due to the induced current of GMM of the vacuum quantum fluctuations \cite{shen02}. Similarly, the gravitomagnetic field
produced by the GMM of the vacuum quantum fluctuations is exactly cancelled by the gravitomagnetic field due to the induced current of the gravitoelectric charge (ordinary mass current) of the vacuum quantum fluctuations \cite{shen02}. In fact, the GMM primordial black holes (PBHs) could be formed due to the energy density fluctuations in the radiation-dominated era in the very early Universe. This is similar to the mechanism of forming an ordinary PBH with the ordinary mass. As Hawking mentioned that a sufficient concentration of the electromagnetic radiation (which is also an example of the `source of pure angular momentum') can cause the gravitational collapse \cite{haw71}, the similar mechanism could form the GMM PBHs, with or without ordinary mass. Specifically, the high densities of `a linear source of pure angular momentum' of GMM and heterogeneous conditions could have led sufficiently dense regions of the GMM to undergo gravitational collapse, forming GMM PBHs in the early Universe. We may  presume that some specific null dust solutions \cite{ste} could be physically described as the `source of pure angular momentum' field or the Taub-NUT spacetime. For example, it might be like the Bonnor beam \cite{bonl} of GMM, or, a beam of neutrinos, surrounded by a vacuum region. The original Bonnor beam \cite{bonl} models an infinitely long, straight beam of light\footnote{Although an infinitely long, straight beam of light implies `a linear source of pure angular momentum', it cannot be relevant for a Taub-NUT spacetime. This is because the Taub-NUT metric is the stationary and spherically symmetric vacuum solution of the Einstein equation, whereas the Bonnor beam of light is the electrovacuum solution of the same. Therefore, the Bonnor beam of GMM should be unique, and could not be the beam of light.} which is gravitationally stable.

\section{Evaporation of spinning PBHs\label{s3}}
Considering the regular matter, Hawking showed \cite{haw71, haw} that the black holes (BHs) create and emit particles as if they are hot bodies with a non-zero temparature. This thermal emission leads to a slow decrement of the ordinary mass of BH and its eventual disappearance. Later,
in a series of papers, Page \cite{page1, page2} calculated the particle emission rates from a spinning BH and showed  that the angular momentum is emitted several times faster than the (ordinary mass-)energy. Therefore, a rapidly spinning Kerr BH should slow down to a nearly non-spinning state before its entire ordinary mass evaporates. Page \cite{page2} showed that a non-spinning PBH with an initial ordinary mass of $\sim 5 \times 10^{11}$ kg would have just decayed away due to the Hawking radiation. Similarly, an initially maximally spinning PBH of ordinary mass $\sim 7 \times 10^{11}$ kg would have just evaporated \cite{page2}.

Now, in case of a Kerr BH, the spin is fully associated with the ordinary mass, and hence the spin cannot exist without the ordinary mass. But, this is not true for Taub-NUT BHs. In this case, the specific angular momentum related to the GMM, which is fundamentally different from the specific angular momentum of a Kerr BH, does not depend on the ordinary mass. Thus, this specific angular momentum or GMM can exist, and can have its corresponding energy, even without the presence of ordinary mass or regular matter. In this paper, we study the role of such GMM for the evaporation of PBHs. For this purpose, we consider non-spinning (the Kerr parameter $a=0$) Taub-NUT BHs with non-zero ordinary mass and GMM values.

\section{\label{sechr}Evaporation of Taub-NUT PBHs\label{s4}}

Here, we investigate the plausible evaporation of Taub-NUT PBHs in a simple way. The mathematical calculation related to the evaporation of a Taub-NUT PBH and a pure GMM PBH are given in Appendix \ref{app1} and Appendix \ref{app2}, respectively.

With the additional contribution from the Misner strings, Hawking et al. \cite{hhp} reproduced the entropy calculated from the action by the usual thermodynamic relations, and showed that the entropy is not just equal to the quarter of the area of bolt, as this is for BHs \cite{hh}. Later, Kerner et al. \cite{km} calculated the Hawking temperature for a special subclass of hyperbolic Taub-NUT-AdS solutions. However, the instability of the Taub-NUT spacetime at the linear level against generic perturbations may be responsible for so far unsuccessful attempts to define a reasonable thermodynamics \cite{holz} for the Taub-NUT spacetime. This may be the reason why the Hawking radiation for this case has not been studied yet in a direct way, despite the possibility to formulate later \cite{hen, bor1} a consistent thermodynamics in the Taub-NUT spacetime.    

In a very recent paper, Foo et al. \cite{foo} proposed to study the Hawking radiation in a different way, which is somewhat similar to the Unruh effect \cite{unruh}. They \cite{foo} investigated an accelerated boundary correspondence which mimics the outgoing Hawking radiation produced by a general class of Taub-NUT BHs. It has been shown \cite{foo} that if a BH contains the GMM $(n)$ along with the regular matter, the particle production is suppressed by the GMM. Considering the particle spectrum of the frequencies of the outgoing and incoming modes $\o$ and $\o'$ respectively, Foo et al. \cite{foo} showed that the particle number decays to zero for large values of $n$ (i.e., $n$ close to $1$ in $G=c=M=1$ unit) for both the regimes: $\o \sim \o'$ and $\o << \o'$. The particle number also vanishes in the limit of small $n$ (i.e., $n$ close to $0$ in  $G=c=M=1$ unit). For the intermediate values of $n$, the particle number grows, the peak value of which depends on the ordinary mass ($M$) of BH. Fig. A5 (a) of \cite{foo} exhibits that the particle number in the peak value is proportional to the ordinary mass $M$ of BH. Thus, if the ordinary mass decreases, the particle number in the peak also decreases. Overall, this interesting research \cite{foo} reveals that the particle number decays to zero depending on various parameter values, and, most importantly, $n$ is responsible for that. Now, as the Misner strings are transparent for geodesics, there seems no opportunity for them to hide
any information and store entropy \cite{bor1}. Thus, no entropy is associated to the Misner string \cite{bor1}, and, one could not
expect the Hawking radiation from the Misner string. Note that, the Misner string is solely associated with the presence of $n$. If $n$ is absent, the Taub-NUT spacetime reduces to the Schwarzchild spacetime which may Hawking radiate, as Hawking proposed. Therefore, one can conclude from the above discussion that the presence of a small $n$ can suppress the Hawking radiation of $M$ too in case of a Taub-NUT PBH, and the evaporation rate of such a PBH could be decayed to zero, depending on the various parameter values. Hence, Taub-NUT PBHs could exist at present even if their initial ordinary mass(-energy) were less than or about $5\times10^{11}$ kg.

Conforming with the standard scenario, one may consider that a Taub-NUT PBH of a sufficiently low initial (ordinary) mass may have been fully Hawking radiated. But as discussed above, $n$ may not be completely evaporated.
This can happen if the GMM evaporation rate is negligible compared to the ordinary mass evaporation rate, and one can roughly estimate 
the Hawking temperature (Eq. \ref{TH} of Appendix \ref{app1}) and the evaporation time (Eq. \ref{tev} of Appendix \ref{app1}) of a Taub-NUT PBH. This scenario does not appear to violate the conclusion drawn in Foo et al. \cite{foo}, and hence it may be reasonable to consider that,
after the evaporation of the entire mass $M$ in time $t_{\rm ev}$ (see Eq. \ref{tev} of Appendix \ref{app1}), an almost unchanged value of $n$ of PBH could exist. This remnant PBH with a GMM $n$ is a pure GMM PBH. Note that such a Taub-NUT solution with $M=0$ is perfectly well-defined \cite{rs2} (also see, for example, appendix to \cite{rs}).

If the GMM evaporates like the ordinary matter, one can calculate the Hawking temperature (Eq. \ref{THm0} of Appendix \ref{app2}), evaporation time (Eq. \ref{tevm0} of Appendix \ref{app2}), etc. following the same procedure to derive the Hawking radiation for an ordinary PBH formed with the regular matter. Comparing Eq. (\ref{tevem}) of Appendix \ref{app1} and Eq. (\ref{teven}) of Appendix \ref{app2}, one may not be able to distinguish between the PBHs made with an ordinary mass and a GMM of the same energy $E=E_M=E_n$, as the order of magnitude of the evaporation times can be the same.

\section{Discussion and Conclusions\label{s5}}
Here, we propose that the lower energy PBHs (equivalent to the ordinary mass $M << 5\times10^{11}$ kg) could still exist in our present Universe, if they contain GMM. This is because either pure GMM PBH could have been formed, or while the regular matter could have decayed away due to the Hawking radiation, the evaporation rate of the GMM could be negligibly small.
It may also be possible that, in the presence of the GMM, the ordinary mass of even an $M << 5\times10^{11}$ kg PBH is not entirely evaporated. Therefore, the remnant pure GMM PBHs and/or PBHs having both ordinary mass and GMM with a relatively low energy (equivalent to $2.176 \times 10^{-8}$ kg $< M < 10^{11}$ kg) could still exist.

Note that the Planck unit of GMM is $n_P=\sqrt{\hbar c^3/G}=6.525$ kg.m.s$^{-1}$. Thus, the available PBHs with the minimum GMM could be $6.525$ kg.m.s$^{-1}$, which is equivalent to the ordinary mass(-energy) $M_P=\sqrt{\hbar c/G}=2.176 \times 10^{-8}$ kg, i.e., the Planck (ordinary) mass. The radius of the horizon of a PBH with Planck (ordinary) mass is $r_h=2GM/c^2=3.23 \times 10^{-35}$ m, whereas for a PBH with Planck GMM is $r_h=Gn/c^3=1.615 \times 10^{-35}$ m. As the energies of both the Planck relics, having the ordinary mass or the GMM, are same, one cannot distinguish between them in principle.
However, the detection of  PBHs with energy equivalent to $M < 5\times10^{11}$ kg could suggest the existence of GMM. Note that the Hawking radiation of such low GMM PBHs are expected to be negligible (otherwise they would have entirely evaporated), and hence they should interact only through the gravitational effect. As the GMM should have the  gravitational effect, the GMM PBHs could contribute to the dark matter.

Low-energy PBHs being constituents of the dark matter, at least partially, is not generally ruled out by observations. If their ordinary mass (but not GMM) has been evaporated over the cosmological time, then for an initial ordinary mass in the range of $\sim 10^6-10^{12}$~kg, $\beta$ may be less than $10^{-18}$ \cite{carr}. Here, $\beta$ is the fraction of the Universe collapsed to form PBHs. But this constraint may not be applicable for a lower initial ordinary mass, and/or if the ordinary mass did not evaporate substantially.
Such plausible low-energy PBHs could have observable effects, such as a seismic signature of PBHs passing through the Sun and other stars \cite{kes}. Note that low-energy PBHs are expected to pass through a star more frequently, and while the signal would be weaker than that for more massive PBHs, such a signal could be detected with future instruments and techniques.
\\

{\bf Acknowledgements :} 
CC thanks Jian Qi Shen and Parthasarathi Majumdar for useful discussions regarding the gravitomagnetic monopole/NUT charge.

\appendix

\section{\label{app1} A rough estimation of Hawking radiation from Taub-NUT black holes
\protect\footnote{Note that this is not the exact Hawking radiation estimation for Taub-NUT BHs, which has so far been unsuccessful (see Sec. \ref{sechr} for details).}}

A Taub-NUT BH has two parameters: ordinary mass ($M$) and GMM or NUT charge ($n$). In this paper, we use the S.I. unit instead of the geometrized unit ($G=c=1$) unless it is specifically mentioned in some places. The S.I. unit of $n$ (angular momentum per unit length \cite{bon}) is ``kg.m.s$^{-1}$''. To convert it to the geometrized unit, one has to multiply $n$ with $G/c^3$, and its unit reduces to the unit of length: meter.
Therefore, the location of the outer horizon ($r_h$) is \cite{cm}
\begin{eqnarray}
 r_h &=& \f{GM}{c^2}+\sqrt{\left(\f{GM}{c^2}\right)^2+\left(\f{Gn}{c^3}\right)^2}
 \\
 &=& \f{G}{c^2} \left[M+\sqrt{M^2+n^2/c^2} \right].
 \label{rh}
\end{eqnarray}

The area $(A)$ of a Taub-NUT BH is expressed as
\begin{eqnarray}
 A=4\pi\left[r_h^2+\left(\f{Gn}{c^3}\right)^2\right]= \f{8\pi G}{c^2} r_h \sqrt{M^2 + n^2/c^2} .
 \label{A}
\end{eqnarray}
The Hawking temperature ($T_H$) is calculated as
\begin{eqnarray}
 T_H=\f{\hbar c}{4\pi k_B r_h}
 \label{TH}
\end{eqnarray}
and, the Hawking luminosity $(L_H)$ is
\begin{eqnarray}
 L_H=\s AT^4=\f{\hbar c^6}{1920\pi G^2}.\f{\sqrt{M^2+n^2/c^2}}{\left(M+\sqrt{M^2+n^2/c^2}\right)^3}
 \label{LH}
\end{eqnarray}
where $\s=\f{\pi^2 k_B^4}{60\hbar^3 c^2}$ is a constant\footnote{$G$ is the gravitational constant, $\hbar$ is the Planck constant, $k_B$ is the Boltzmann constant and $c$ is the speed of light in vacuum.}.

If the evaporation rate of the GMM is negligible compared to the ordinary mass evaporation rate (see Sec. \ref{sechr}), we can write
\begin{eqnarray}
 \f{d(n/c)}{dt} <<\f{dM}{dt}.
 \label{as}
\end{eqnarray}

If the energy ($E$) dissipates due to the Hawking radiation,  one obtains the following from Eq. (\ref{LH}) and Eq. (\ref{as}):
\begin{eqnarray}
 -c^2\f{dM}{dt}=\f{\hbar c^6}{1920\pi G^2}.\f{\sqrt{M^2+n^2/c^2}}{\left(M+\sqrt{M^2+n^2/c^2}\right)^3}.
\end{eqnarray} 
Integrating the above equation, we obtain the ordinary mass evaporation time ($t_{\rm ev}$) as:
\begin{eqnarray} \nonumber
\int_0^{t_{\rm ev}} dt &=& - \f{1920\pi G^2}{\hbar c^4} \int_M^0 \f{\left(M+\sqrt{M^2+n^2/c^2}\right)^3}{\sqrt{M^2+n^2/c^2}} dM
 \\
 \Rightarrow t_{\rm ev} &=& \f{640\pi G^2}{\hbar c^7} \nonumber\left[4M^3c^3+3Mn^2 c-n^3 \right.
 \\
 && \left. + c(4M^2c^2+n^2) \sqrt{M^2+n^2/c^2} \right] .
 \label{tev}
\end{eqnarray}
For $n=0$, Eq. (\ref{tev}) reduces to
\begin{eqnarray}
 t_{\rm ev}\big|_{n=0}=\f{5120\pi G^2}{\hbar c^4}M^3,
 \label{tevn0}
\end{eqnarray}
which is the well-known expression of the Hawking radiation for the evaporation of a non-spinning BH. In terms of the energy (i.e., $E_M \sim Mc^2$), Eq. (\ref{tevn0}) can be written as 
\begin{eqnarray}
 t_{\rm ev}\big|_{n=0} \sim \f{5120\pi G^2}{\hbar c^{10}}E_M^3 .
 \label{tevem}
\end{eqnarray}

\section{\label{app2}{Evaporation of a pure gravitomagnetic monopole black hole}}

Using Eqs. (\ref{rh}--\ref{LH}) we obtain the radius of the horizon of a pure GMM black hole as:
\begin{eqnarray}
 r_{ hn}=\f{Gn}{c^3},
\end{eqnarray}
area:
\begin{eqnarray}
 A_n= \f{8\pi G^2 n^2}{c^6},
\end{eqnarray}
the Hawking temperature:
\begin{eqnarray}
 T_{{\rm H}n}=\f{\hbar c^4}{4\pi k_B G n}
 \label{THm0}
\end{eqnarray}
and, the Hawking luminosity
\begin{eqnarray}
 L_{{\rm H}n}=\s AT^4=\f{\hbar c^8}{1920\pi G^2 n^2} .
\end{eqnarray}
As the energy of the GMM is $E_n \sim nc$, we obtain
\begin{eqnarray}
 -c\f{dn}{dt}=\f{\hbar c^8}{1920\pi G^2 n^2}.
 \label{dndt}
\end{eqnarray} 
 Solving Eq. (\ref{dndt}) we obtain the evaporation time ($t_{{\rm ev}n}$) of a Taub-NUT BH with GMM as,
 \begin{eqnarray}
 \int_0^{t_{{\rm ev}n}} dt &=& - \f{1920\pi G^2}{\hbar c^7} \int_n^0 n^2 dn
 \\
 \Rightarrow t_{{\rm ev}n} &=& \f{640\pi G^2}{\hbar c^7}n^3 ,
 \label{tevm0}
\end{eqnarray} 
in case, the evaporation of $n$ is permitted.
In terms of the energy (i.e., $E_n \sim nc$), Eq. (\ref{tevm0}) can be written as 
 \begin{eqnarray}
 t_{{\rm ev}n} \sim \f{640\pi G^2}{\hbar c^{10}}E_n^3 .
 \label{teven}
\end{eqnarray}


\begin{thebibliography}{}


\bibitem{shen02} J. Q. Shen, Gen. Rel. Gravit. {\bf 34}, 1423 (2002).

\bibitem{shen04} J. Q. Shen, Ann. Phys. (Leipzig) {\bf 13}, 532 (2004).

\bibitem{rs} S. Ramaswamy, A. Sen, J. Math. Phys. (N.Y.) {\bf 22}, 2612 (1981).

\bibitem{str} N. Straumann, General Relativity with applications to Astrophysics, Springer, Berlin (2009).

\bibitem{nut} E. Newman, L. Tamburino, T. Unti, J. Math. Phys. {\bf 4}, 915 (1963).

\bibitem{bon} W. B. Bonnor, Proc. Camb. Phil. Soc. {\bf 66}, 145 (1969).

\bibitem{dow} J. S. Dowker, Gen. Rel. Grav. {\bf 5}, 603 (1974).

\bibitem{hel} S. W. Hawking, G. F. R. Ellis, The large scale structure of space-time, Cambridge University Press (1994).

\bibitem{gcyl} M. Ghasemi-Nodehi, C. Chakraborty, Q. Yu, Y. Lu, Eur. Phys. J. {\bf C 81}, 939 (2021).

\bibitem{mis} C. W. Misner, J. Math. Phys. {\bf 4}, 924 (1963).

\bibitem{rs2} S. Ramaswamy, A. Sen, Phys. Rev. Lett. {\bf 57}, 1088 (1986).

\bibitem{cle1} G. Cl\'ement, D. Gal'tsov, M. Guenouche, Phys. Lett. {\bf B 750}, 591 (2015).

\bibitem{cle2} G. Cl\'ement, D. Gal'tsov, M. Guenouche, Phys. Rev. {\bf D 93}, 024048 (2016).

\bibitem{hen} R. A. Hennigar, D. Kubizn\'ak, R. B. Mann,  Phys. Rev. {\bf D 100}, 064055 (2019).

\bibitem{bor1} A. B. Bordo, F. Gray, D. Kubizn\'ak, J. High Energy Phys. {\bf 2019}, 119 (2019).

 \bibitem{nov} I. D. Novikov, {\it Evolution of the Universe},
Cambridge University Press, New York (1983)

\bibitem{fn} J. Friedman et al., Phys. Rev. {\bf D 42}, 1915 (1990).

\bibitem{lnbl} D. Lynden-Bell, M. Nouri-Zonoz, Rev. Mod. Phys. {\bf 70}, 427 (1998).

 \bibitem{liu} C. Liu, S. Chen, C. Ding, J. Jing, Phys. Lett. {\bf B 701}, 285 (2011).
 
 \bibitem{cc} C. Chakraborty, Eur. Phys. J. {\bf C 75}, 572 (2015). 

\bibitem{kag} V. Kagramanova, J. Kunz, E. Hackmann, C. L\"ammerzahl, Phys. Rev. {\bf D 81}, 124044 (2010)

\bibitem{cbgm} C. Chakraborty, S. Bhattacharyya, Phys. Rev. {\bf D 98}, 043021 (2018).

\bibitem{cbgm2} C. Chakraborty, S. Bhattacharyya, JCAP {\bf 05} (2019) 034.

\bibitem{shenp} Personal email communication with J. Q. Shen.

\bibitem{bon60} W. B. Bonnor, Proc. Phys. Soc. {\bf 76}, 891 (1960).

\bibitem{haw71} S. W. Hawking, MNRAS {\bf 152}, 75 (1971).

\bibitem{ste} H. Stephani et al.,  Exact Solutions of Einstein's Field Equations, Cambridge: Cambridge University Press (2003).

\bibitem{bonl} W. B. Bonnor, Comm. Math. Phys. {\bf 13}, 163 (1969).

\bibitem{haw} S. W. Hawking, Commun. Math. Phys. {\bf 43}, 199 (1975).

\bibitem{page1} D. N. Page, Phys. Rev. {\bf D 13}, 198 (1976).

\bibitem{page2} D. N. Page, Phys. Rev. {\bf D 14}, 3260 (1976).

 \bibitem{hhp} S. W. Hawking, C. J. Hunter, D. N. Page, Phys. Rev. D 59, 044033 (1999).

 \bibitem{hh} S. W. Hawking, C. J. Hunter, Phys. Rev. D 59, 044025 (1999).
 
 \bibitem{km} R. Kerner, R. B. Mann, Phys. Rev. D 73, 104010 (2006).
 
 \bibitem{holz} G. Holzegel, Class. Quantum Grav. 23 (2006) 3951.

\bibitem{foo} J. Foo, M. R. R. Good, R. B. Mann, Universe {\bf 2021}, 7, 350.  

\bibitem{unruh} W. G. Unruh, Phys. Rev. D 14, 870 (1976).
 
\bibitem{carr} B. Carr et al., Rep. Prog. Phys. {\bf 84}, 116902 (2021).

 \bibitem{kes} M. Kesden 1, S. Hanasoge, Phys. Rev. Lett. {\bf 107}, 111101 (2011).
   
  \bibitem{cm} C. Chakraborty, P. Majumdar, Class. Quantum Grav. {\bf 31}, 075006 (2014).

\end{thebibliography}
\end{document}